\documentstyle[12pt,aasms]{article}
\tighten
\eqsecnum

\slugcomment{ draft \today}

\begin{document}

\font\twelvei = cmmi10 scaled\magstep1 
       \font\teni = cmmi10 \font\seveni = cmmi7
\font\mbf = cmmib10 scaled\magstep1
       \font\mbfs = cmmib10 \font\mbfss = cmmib10 scaled 833
\font\msybf = cmbsy10 scaled\magstep1
       \font\msybfs = cmbsy10 \font\msybfss = cmbsy10 scaled 833
\textfont1 = \twelvei
       \scriptfont1 = \twelvei \scriptscriptfont1 = \teni
       \def\mit{\fam1 }
\textfont9 = \mbf
       \scriptfont9 = \mbfs \scriptscriptfont9 = \mbfss
       \def\bmit{\fam9 }
\textfont10 = \msybf
       \scriptfont10 = \msybfs \scriptscriptfont10 = \msybfss
       \def\bmsy{\fam10 }

\def\etal{{\it et al.~}}
\def\eg{{\it e.g.}}
\def\ie{{\it i.e.}}
\def\lsim{\raise0.3ex\hbox{$<$}\kern-0.75em{\lower0.65ex\hbox{$\sim$}}} 
\def\gsim{\raise0.3ex\hbox{$>$}\kern-0.75em{\lower0.65ex\hbox{$\sim$}}}
\def \msol {\rm{M}$_\odot$}
\def \msol {\rm{M}$_\odot$}
\def \mdot {\rm{M}$_\odot$~yr$^{-1}$}
\def \lam {$\lambda$}
\def \kms{km~$\rm{s}^{-1}$}
\def \cc{$\rm{cm}^{-3}$}
\def \arcs{\char'175}
\def \lam{$\lambda$}
\def \micra{$\mu$m}

\title{Hydrodynamical Models of Outflow Collimation in YSOs}

\author{Adam Frank \altaffilmark{1}, Garrelt Mellema \altaffilmark{2}}

\altaffiltext{1}{ Hubble Fellow; Department of Astronomy, 
     University of Minnesota,
    Minneapolis, MN 55455; e-mail: afrank@astro.spa.umn.edu}
\altaffiltext{2}{ Stockholm Observatory, S-13336 Saltsj\"obaden,
    Sweden; email: garrelt@astro.su.se }

\clearpage

In this paper we explore the physics of time-dependent hydrodynamic
collimation of jets from Young Stellar Objects (YSOs).  Using
parameters appropriate to YSOs we have carried out high resolution
hydrodynamic simulations modeling the interaction of a central wind
with an environment characterized by a toroidal density distribution
which has a moderate opening angle of $\theta_{\rm \rho} \approx 90^{\rm
o}$ The results show that for all but low values of the equator to
pole density contrast the wind/environment interaction produces
strongly collimated supersonic jets.  The jet is composed of shocked
wind gas.  Using analytical models of wind blown bubble evolution we
show that the scenario studied here should be applicable to YSOs and
can, in principle, initiate collimation on the correct scales ($R \lsim
100$~AU).  Comparison of our simulations with analytical models
demonstrates that the evolution seen in the simulations is a mix of
wind-blown bubble and jet dynamics. The simulations reveal a number of
time-dependent non-linear features not anticipated in previous
analytical studies.  These include: a prolate wind shock; a chimney of
cold swept-up ambient material dragged into the bubble cavity; a plug
of dense material between the jet and bow shocks.  We find that the
collimation of the jet occurs through both de Laval nozzles and
focusing of the wind via the prolate wind shock. Using an analytical
model for shock focusing we demonstrate that a prolate wind shock
can, by itself, produce highly collimated supersonic jets.

Animations from these simulations are available over the internet at
WWW address http://www.msi.umn.edu/Projects/twj/jetcol.html

\keywords{ISM: Jets and Outflows - hydrodynamics - star: formation}

\clearpage

\section{Introduction}
The propagation of jets associated with Young Stellar Objects (YSOs)
has been well studied both analytically (\cite{RaKof92}) and with
sophisticated numerical tools (\cite{BlFrKo90}, \cite{StoNor94a}).
These theoretical investigations, in conjunction with the growing
data-base of high-resolution observations, have been extremely useful
in understanding the hydrodynamics of HH jets and HH objects.  The
origin of these jets, however, remains an issue which has yet to be
resolved.  The obscuring dust and gas surrounding young stars has made
it difficult to observationally determine physical conditions which
can constrain collimation models for either the HH jets or the
(perhaps related) bipolar CO outflows.  In the absence of these
constraints a number of collimation mechanisms have been
proposed which, broadly speaking, fall into two categories: pure
hydrodynamic models and magnetohydrodynamic models.

Many of the purely hydrodynamic studies produce well collimated
supersonic outflows by invoking de Laval nozzles (\cite{Kon82},
\cite{RaCan89}).  In these models an initially spherical stellar wind
interacts with the surrounding medium and is shocked producing a high
temperature cavity. If the walls of the cavity take on the appropriate
``nozzle'' configuration, transsonic solutions for the flow exist
leading to the formation of a supersonic jet.  There are, however, a
number of problems with the de Laval nozzle scenarios.  The nozzles in
the cavity may be unstable (\cite{KoMc92}) and the high densities in
the shocked gas may produce cooling distances to short to allow a
``hot bubble'' to form within the cavity (\cite{PelPud92}).

Another class of hydrodynamic collimation models which rely on the
other extreme of cooling length scales was explored by Cant\'o (1980),
Cant\'o \& Rodriguez (1983) and Cant\'o, Tenorio-Tagle \& Rozyczka (1988).
In these models strong radiative losses create a thin aspherical shell.  
After the freely expanding wind strikes
the shock at an oblique angle it is redirected to flow along the walls of 
the shell. At the vertex of the aspherical (prolate) shell a converging
conical flow is established which produces a jet. The main problem with
these models is the size scale of the steady-state shell (based on
achieving pressure equilibrium) which is larger than the size of observed 
collimation regions ($R \approx 100$ AU, see \cite{BurStp95}). 

The strong evidence supporting circumstellar disks (\cite{Strom94})
and magnetic fields around T Tauri stars has led to a different set of
scenarios for producing well collimated jets. In these MHD
``disk-wind'' models the outflows are centrifugally driven by a
magnetized accretion disk.  Considerable effort has gone into the 
theory that the magnetic
field in the disk forces the accreting gas into co-rotation.
Centrifugal acceleration and magnetic force then lift the gas off the
disk producing a wind which is eventually collimated into a jet as the
field lines bend back towards the disk/star rotation axis
(\cite{Kon89,Pud91}). The details of the gas acceleration depend on
the magnetic field configuration (\cite{PelPud92}, \cite{Todo93},
\cite{WarKon93}, \cite{Hollet95}), the star-disk interaction at the
boundary layer (\cite{Cam93}, \cite{NajShu94}), and the stability in
co-rotation (\cite{UchShi85}).  These disk-wind models are very
promising and the current consensus appears to be that the outflow
collimation occurs through some kind MHD process.  However, these
models also have their problems. MHD disk-wind models suffer from
difficulties in producing the correct disk-field orientations
(\cite{Shu91}).  Other magneto-gasdynamic models suffer from
uncertainties in the actual field strengths and orientations achieved
in YSOs (\cite{Bal93,Good90,Hey87}). In addition it is not clear if or
how long the field can maintain the focusing of the flow into a
tightly collimated jet or if the collimation can be achieved on the
correct scales (\cite{StoNor94b}).

To further complicate matters there is now increasing observational
evidence that that collimated YSO flows are {\it essentially}
unsteady. HH jets show signs of velocity variations (\cite{Mors92})
and entrainment (\cite{Harti93}), suggesting that both the driving of
the YSO wind and its interaction with the circumstellar environment
are time-dependent processes.
 
Clearly a great deal of progress remains to be made in our
understanding of the YSO collimation process. Because of the
complexity of both the flows and the underlying physics, numerical
simulations are an effective tool for studying outflow collimation.
While there is an abundance of simulations of fully developed jets
(\cite{BlFrKo90,StoNor94a}) surprisingly little numerical work has
been done on their collimation or on the collimation of the bipolar
outflows (see e.g. \cite{Nor93}, Chernin \& Masson 1994).  In this
paper we seek to re-examine gasdynamical collimation using
high-resolution numerical simulations.  Recent numerical studies of
the formation of Planetary Nebulae have shown that excellent
collimation can be achieved through the interaction of a
spherical central wind with a toroidal circumstellar environment
(\cite{Melal91}, \cite{IcBaFr92}, \cite{Ickeal92}).  These studies have
demonstrated that nonlinear and time-dependent gasdynamic effects
provide collimating mechanisms which were unanticipated or not fully
appreciated in previous analytical models.  The time-dependent
collimation processes have been named 'Shock-Focused Inertial
Confinement' (SFIC).  In the SFIC mechanism the interaction of the
inertia of a toroidal environment with the thermal pressure of the
shocked wind produces a well collimated jet.
 
In a preliminary study \cite{FrNor94} (hereafter FN94) investigated
the SFIC mechanism in the context of YSOs. FN94 used a toroidal
density environment which included an accretion flow.  Their
simulations showed that focusing at the inner shock could produce
strongly collimated supersonic flows.  While these results are
promising a deeper understanding of the SFIC mechanism is required
before a serious application to YSO jets can be attempted.  In this
paper we attempt to take some steps in this direction by investigating
an idealized adiabatic version of SFIC collimation.  Using numerical
simulations and analytical estimates we will attempt to further
identify the basic processes at work in SFIC collimation and put
limits on their the applicability to YSOs.
 
In re-examining hydrodynamic collimation our intent is not to try to
push MHD models aside.  There are many reasons for expecting
magnetic fields to be important in producing at least some jets,
particularly those associated with T Tauri stars. But a robust model
of hydrodynamic collimation could be used to produce jets even in
those cases where the MHD scenarios such as the disk-winds models can
produce only poorly collimated winds or where MHD collimation is not
effective on all scales relative to the full length of the jet. We
note that the hydrodynamics we explore in this paper has elements
similar to both Cant\'o's 1980 model and de Laval nozzles.  Thus
we intend to study the SFIC mechanism as a general set of processes
which may individually operate in some form across a variety of length
scales rather than as the definitive model for the production of YSO
jets. We note here that since we are exploring
the effect of the circum-protostellar
environment on jet formation our models may be particularly relevent to the
more deeply embedded class 0 objects.

The organization of the paper is as follows: In section II we describe
of the numerical method and initial conditions used in our
simulations.  In Section III we provide some analytical estimates of
the range of applicability of the SFIC mechanism with respect to
initial conditions. In section IV we examine the results of our
numerical models.  In section V we explore the collimation mechanisms
seen in the simulations.  Finally in section VI we present our
conclusions along with a discussion of some issues raised by the
simulations.

\section{Numerical Method and Initial Conditions}
The numerical model (initial conditions and governing equations) we
have constructed for our simulations capture the essential
characteristics of the environment we wish to study: a central wind
interacting with a toroidal environment.  Our ultimate goal is to
investigate the collimation of realistic YSO jets through the SFIC
mechanism described in section I.  But in this paper, we focus on the
physics of SFIC collimation in an idealized environment. We state
explicitly that the initial conditions used here are not meant to be
realistic in the sense of modeling an actual YSO environment.  There
are a number of scenarios for gravitational collapse that lead to star
formation: collapse of a rotating spherical cloud (\cite{TSC84});
collapse of a spherical cloud threaded by an ordered magnetic field
(\cite{GalShu93}); collapse of a flattened filament (\cite{Hartet94}).
Each of the collapse scenarios listed above would produce a toroidal
density distribution. However, the explicit form of those
distributions as well as the form of the velocity fields they create
would vary considerably from one scenario to the next.  Since we
intend to study the SFIC mechanism in the context of specific collapse
scenarios in a future work, we use here their common characteristics
as initial conditions for the environment.

The gasdynamic interactions we wish to study are governed by the Euler
equations.  In our numerical model we express these in azimuthally
symmetric cylindrical coordinates:
\begin{equation}
{\partial\rho \over \partial t} + 
{1 \over r}{\partial \rho u_{\rm r} \over \partial r} +
{\partial \rho u_{\rm z} \over \partial z} =
0, 
\label{masscon}
\end{equation}
\begin{equation}
{\partial\rho u_{\rm r} \over \partial t} + 
{1 \over r}{\partial \rho {u_{\rm r}}^2 \over \partial r} +
{\partial \rho u_{\rm r} u_{\rm z} \over \partial z} =
 - {\partial p \over \partial r} 
\label{conmomr}
\end{equation}
\begin{equation}
{\partial\rho u_{\rm z} \over \partial t} + 
{1 \over r}{\partial \rho u_{\rm z} u_{\rm r} \over \partial r} +
{\partial \rho {u_{\rm r}}^2 \over \partial z} =
 - {\partial p \over \partial z}, 
\label{conmomz}
\end{equation}
\begin{equation}
{\partial E \over \partial t} + 
{1 \over r}{\partial (E + p) u_{\rm r} \over \partial r} +
{\partial (E + p) u_{\rm z} \over \partial z} = 0, 
\label{conmome}
\end{equation}
and
\begin{equation}
E = {1 \over 2} \rho ({u_{\rm r}}^2 + {u_{\rm z}}^2) + 
{1 \over (\gamma - 1)}p, 
\label{edef}
\end{equation}
where the terms have their usual meaning. To solve these equations we
use the Total Variation Diminishing (TVD) method of \cite{harten83} as
implemented by \cite{Ryuet95}.  TVD is an explicit method for solving
hyperbolic systems of equations.  It achieves second order accuracy by
finding approximate solutions to the Riemann problem at each grid
boundary while remaining non-oscillatory through the application of a
lower order monotone scheme. The implementation of the TVD method used
here is robust and requires even less CPU time than older methods such
as the Flux Corrected Transport (FCT) schemes (Boris \& Book 1973).

Note that equations \ref{masscon} through \ref{edef} do not include
the effects of rotation and gravitational fields.  We also leave out
the effects of radiative energy losses.  As said at the beginning of
this section, we focus here on the simplest case, the purely
adiabatic, purely gas dynamical collimation of jets. As we shall
demonstrate, even under these constraints the flow pattern which
develops is quite complex.  We feel it is important to understand the
dynamics of these flows before adding additional physics.
Below we provide some justification for ignoring the effects of
gravity and rotation and in section 3 we discuss the potential
role of radiative cooling in our collimation mechanisms.

The most important feature of the environment for this study is the
presence of a initial density contrast between pole and equator.
Because gravity and rotation can both determine the shape of the
density distribution we will use a toroidal density distribution with
a radial power law appropriate to an in-falling cloud in the central
potential of a protostar.  The initial density distribution we chose
to work with takes the following form:
\begin{equation}
\rho(R,\theta) = {{\dot M}_a \over 4\pi R^2} ({2GM \over R})^{- {1 \over 2}}
{\lbrace 1 - {\zeta \over 6}{\lbrack 13P_{\rm 2} ( \cos (\theta)) -1
\rbrack}\rbrace}
\label{rhoin}
\end{equation}
Note that eq \ref{rhoin} is expressed in spherical coordinates.  In
the rest of this paper we will use $R$ and $r$ to denote the spherical
and cylindrical radii respectively where $R = \sqrt{r^2 + z^2}$ and
$\theta = \tan^{-1}(r/z)$.  In eq \ref{rhoin} ${\dot M}_{\rm a}$ is
the accretion mass loss rate and M is the mass of the star.  Equation
\ref{rhoin} is a modified form of eq 96 from \cite{TSC84} (originally
derived by \cite{Ulrich76}).  We use it here because produces the
required toroidal geometry as well having the $R^{\rm -{3 \over 2}}$
radial dependence, appropriate to a freely falling envelope.  The
parameter $\zeta$, which determines the flattening of the cloud, is
normally a function of radius (due to conservation of angular
momentum).  Since one of the principal goals of this study is to
isolate the effect of the equator to pole density contrast ($q =
{\rho_{\rm e} / \rho_{\rm p}}$) on the SFIC collimation process we
have modified the original equation making $\zeta$ constant and
treating it as an input parameter.  Also in collapse schemes like that
described \cite{Hartet94} in which there is no rotation the pole to
equator density contrast will not be a strong function of radius.  The
relation between $\zeta$ and $q$ is
\begin{equation}
q = {12 + 15\zeta \over 12 - 24\zeta}.
\label{qdef}
\end{equation}

In the present application we set the velocity in the environment
equal to zero to allow comparison with analytical
predictions.  In reality the cloud will be falling inward at
velocities on the order of $v_{\rm g} \approx \sqrt{GM/R}$.  
Gravity and the infall motions of the
cloud are, of course, the most important physical components 
that actually form the star.
However, we ignore these aspects of the problem in the present study
because the outward velocities
produced by the wind/environment interaction are much greater $v_{\rm
g}$ and our goal is 
understanding the more restricted problem
of hydrodynamic jet collimation dynamics.
In this study we are more interested in the formation of the jets than
in the formation of the star. 
Thus we can ignore the dynamical effect of the infall velocity.  
Although the condition $v_o >> v_{\rm g}$ does not hold for 
all shock positions (as
we will see the equatorial shock moves out very slowly), we found in
tests that neglecting the accretion velocity does not make a substantial
difference for the bubble structure. A similar argument can be made
with respect to rotational velocities since infalling material must
rotate at speeds less than the Keplerian value.

In Fig~1 we present a contour plot of the density distribution given
by eq \ref{rhoin}.  Superimposed on top of the density contours
we plot a line marking the full-width half maximum of the
distribution, i.e. the angle at which the density falls by half its
value at the equator.  Note that the ``opening angle'' of the torus
defined in this way is $\sim 90^{\rm o}$. Thus while our initial
conditions do not describe a thin disk they certainly do not
correspond to ``funnels'' either and one would not, {\it a priori},
expect them to produce collimated jets.

We assume that the temperature in the cloud is constant.  The ambient
pressure is then set by the equation of state for an ideal gas.  Note
that this implies an outward pressure gradient in the environment.
But the low temperatures in the environment ($T < 500$~K) ensure
small values for the sound speed and we do not see appreciable
evolution of the environment during a simulation.

The initial conditions for the spherically symmetric central wind are
fully specified by its mass loss rate ${\dot M}_{\rm w}$, velocity
$V_{\rm w}$, and temperature $T_{\rm w}$.  In the simulations the wind
is fixed in a spherical region ($R < R_{\rm o}$) at the center of the
grid.  The use of cylindrical ($r,z$) coordinates prohibits exact
specification of the inner wind ``sphere'' and produces a
staircase-like pattern.  This leads to small oscillations in the
temperature and velocity of the freely flowing wind.  To check that
these effects do not degrade the computed solutions we have compared
our simulations of both spherical and aspherical wind blown bubbles
with analytical models (\cite{Wevet77}, \cite{KoMc92}) as well as
simulations computed with other numerical methods in spherical
coordinates (\cite{Frank92}).  We find no appreciable effects of the
imperfect inner wind boundary on the bubble dynamics. The numerical
simulations reproduced the self-similar analytical models usually to
better than 10\%, but never worse than 15\% in terms of shock radii
and velocities.

We have run over 30 simulations exploring a variety of initial
parameters. In this paper we present the results of eight of
these. Their input parameters are listed in Table 1.

\section{Applicability to YSOs}
The SFIC mechanism produces jets inside wind-blown bubbles with the
well-known `three shock' structure. The first shock faces outward into
the environment and is often called the ``outer'' shock.  Following
\cite{KoMc92} we refer to this structure as the ``ambient shock''
($R_{\rm s}$).  The second shock faces into the central wind.  It is
responsible for decelerating and heating the wind.  In the literature
it is sometimes referred to as the ``inner'' or ``reverse'' shock.  We
call it the ``wind shock'' ($R_{\rm sw}$).  Between these two shocks
is the contact discontinuity (CD) separating the shocked wind gas from
the shocked ambient gas.

Previous simulations of SFIC collimation have shown that ``flow
focusing'' at the wind shock is an important process for collimating SFIC
jets.  When the spherical wind strikes the aspherical wind shock at an
oblique angle it is partly redirected into a beam aligned to the
poles of the toroidal density distribution.  Since the focusing occurs
{\it inside} the wind-blown bubble the wind and ambient shocks must be
well separated if the SFIC mechanism is to be effective. A similar
argument holds for the presence of de Laval nozzles where the shocked
wind fills a cavity that acts as a reservoir of thermal energy to be
converted into bulk kinetic energy of a jet.  Thus a necessary
condition for the hydrodynamic collimation mechanisms studied here to
be effective is that $R_{\rm s} >> R_{\rm sw}$ over at least the
polar sector of the wind-blown bubble. Using simple analytical
estimates we demonstrate below that it is possible to obtain this
condition and, in principle, achieve hydrodynamic collimation through
either/both de Laval nozzles and/or the SFIC mechanism on scales that
are consistent with YSO observations.

A shock wave is called radiative if the cooling time for the
post-shock gas is shorter than the dynamical time scale for its
evolution.  If both the wind and ambient shocks are fully radiative
then their separation will be small.  Thermal energy gained at the
shocks is quickly radiated away and, lacking pressure support, the
shocked wind and shocked ambient material collapse on to each other
forming a thin dense shell.  The bubble will then be driven by
the momentum of the wind.  Given a sufficient equator to pole density
contrast in the ambient medium, a momentum driven bubble
will become highly aspherical.  Just as in the SFIC mechanism
described above, the wind in an aspherical radiative bubble will
strike the inner shock at an oblique angle causing focusing towards
the poles.  In this situation, however, the shocked wind material must
slide along the inner edge of the thin shell and can only form a jet
directly over the poles were its conical stream converges.  Such a
configuration was the basis of Cant\'o's 1980 model for the production
of HH objects (see also \cite{TCR88}). Numerical simulations of the
SFIC mechanism with radiative cooling included also produce this type
of flow pattern (\cite{MelFr96a}, \cite{MelFr96b}).

When the cooling time for the shocked wind $t_{\rm c}$ is comparable
to, or longer than, the dynamical time scale for the bubble's
evolution $t_{\rm d}$, the post-shock wind does not lose its thermal
energy to radiation and has enough pressure to push the wind and
ambient shocks apart.  A hot cavity of shocked wind material forms,
filling a large fraction of the bubble's volume.  The swept-up shocked
ambient material, however, remains confined to relatively thin shell.
This is the domain where collimation through de Laval nozzles and the
SFIC mechanism is possible. The bubble is said to be ``energy
conserving'' or ``adiabatic''.

If we wish to determine where in YSO parameter space SFIC jets might
form, we must determine the radii at which the shocks in a wind-blown
bubble begin to separate.  The usual means of doing this is to
determine the radius at which a bubble makes a transition from being
radiative (or momentum driven) to being adiabatic (or energy driven).
There is, however, another possibility.

In their work on the evolution of wind blown bubbles \cite{KoMc92}
showed that between the radiative and adiabatic configurations lies
another evolutionary state which they called the Partially Radiative
Bubble (PRB).  In a PRB the cooling time for the gas is shorter than
the age of the bubble but longer than the time it takes for the
unshocked wind to reach the wind shock i.e. $t_{\rm cross} < t_{\rm cool}
< t_{\rm d}$ where $t_{\rm cross} = R_{\rm s}/v_{\rm w}$.  While at
any time in the PRB stage most of the shocked wind will have cooled,
the wind material which has recently past through the shock will still
be hot enough to keep $R_{\rm s} >> R_{\rm sw}$. Thus the appropriate
transition radius we must find is $R_{\rm t}({\rm PRB})$: the distance
at which a radiative bubble makes the transition to a PRB.

Of course not all bubbles will make a transition to the PRB stage.
But as long as the radial density distribution is such that
\begin{equation} 
\rho(r)=\rho_{\rm 01}r^{\rm -k} ,  
\label{rhdep}
\end{equation}
the range of accretion rates ${\dot M}_{\rm a}$, wind mass loss rates
${\dot M}_{\rm w}$ and wind velocities $v_{\rm w}$ appropriate for
YSOs, is such that all wind-blown bubbles with $k = 3/2$ will enter
the PRB phase. Below we calculate the PRB transition radius $R_{\rm
t}({\rm PRB})$.

The expansion of a radiative, momentum-driven bubble into an
environment given by eq \ref{rhdep} is
\begin{equation}
 R_{\rm s} = \left( {(3-k) \dot{M}_{\rm w}v_{\rm w}\over
12\pi\rho_{\rm 01}} \right)^{\rm 1 \over 4-k} \left({t \over a}\right)^{\rm 2
\over 4-k}\;, \label{rsh1} 
\end{equation} 
where $R_{\rm s}$ is the radius to the outer shock, and
$a=\sqrt{2/(12-3k)}$. The cooling time for post shock gas $t_{\rm c}$
can be estimated from the familiar results of Kahn (1976)
\begin{equation}
t_{\rm c} = {Cv_{\rm s}^3 \over \rho_{\rm pre}}, 
\label{tcool}
\end{equation} 
where $C = 6\times 10^{-35}$~g~cm$^{-6}$~s$^4$ and $\rho_{\rm pre}$ is
the preshock density.  By setting $t_{\rm cool} = t_{\rm cross}$ and
using eqs \ref{rhdep} through \ref{tcool} \cite{KoMc92} derived the
transition time scale
\begin{equation} 
t_{\rm t}({\rm PRB}) = \sqrt{ 
             {\lbrack
             {{1\over 4 \pi C} \rbrack}^{\rm 4-k}
              {12 \pi \rho_{\rm 01} a^2 \over (3 - k)}
                 ~ {\dot M}_{\rm w}}^{\rm (3-k)}
                 {v_{\rm w}}^{\rm -(21 - 5k)} .
                      }
\label{ttprb}
\end{equation}
Substituting this into equation \ref{rsh1} gives
\begin{equation}
R_{\rm t}({\rm PRB}) = ({1\over 4 \pi C}){\dot M}_{\rm w} {v_{\rm w}}^{\rm -5}.
\label{rtprb}
\end{equation}
Note that while $t_{\rm t}({\rm PRB})$ depends on $k$ and $\rho_{\rm
01}$, $R_{\rm t}({\rm PRB})$ does not. Note also the strong dependence of
$R_{\rm t}({\rm PRB})$ on velocity.

For comparison let us also calculate $R_{\rm t}({\rm AD})$, the radius at
which the bubble becomes fully adiabatic.  \cite{KoMc92} calculate
the time scale for a PRB to become an adiabatic bubble but it depends
on $\gamma_{\rm sw}$ the ratio of specific heats in the PRB's shocked
wind.  Since this quantity cannot be calculated in a straight forward
way we will derive $R_{\rm t}({\rm AD})$ by equating the age of a momentum
driven bubble with the cooling time for the post-shock wind $t_{\rm c}
= t_{\rm d}$.  In this case eq \ref{tcool} takes the form
\begin{equation}
t_{\rm c} \approx
	{4\pi Cv_{\rm w}^4 R_s^2 \over \dot{M}_{\rm w}}.
\label{tcols} 
\end{equation}
In eq \ref{tcols} we have assumed that the shock velocity $v_{\rm s}$
is approximately equal to the wind velocity $v_{\rm w}$.  Inverting eq
\ref{rsh1} gives a dynamical time $t_{\rm dyn}$.  Using this we find,
\begin{equation}
  R_{\rm t}({\rm AD}) =\left( a \over 4\pi C\right)^{\rm {2 \over k}} 
            \left( 12 \pi \rho_{01}\dot{M}_{\rm w} \over
                 3-k\right)^{\rm {1 \over k}}
	    v _{\rm w}^{-{9 \over k}}.
\label{rtran}
\end{equation}

In Fig~2 we show $R_{\rm t}({\rm PRB})$ and $R_{\rm t}({\rm AD})$
versus $v_{\rm w}$ for ${\dot M}_{\rm w} = 10^{-7}$ \mdot. The curve
for $R_{\rm t}({\rm AD})$ was calculated using ${\dot M}_{\rm a} =
10^{-6}$ \mdot\ and $k=3/2$.  Figure 2 demonstrates that the bubble
enters the partially radiative stage at $R < 100$~AU for $v_{\rm w} >
150 $ \kms .  The results of \cite{Hirthet94} and recent HST images
(\cite{BurStp95}) give sizes for the collimation region on the order
of $100$~AU.  In addition most HH jets are observed to have velocities
on the order of $200$ \kms\ or more.  In a collimation model that
relies on either shock focusing or de Laval nozzles these jet speeds
imply even higher wind speeds.  Thus Fig~2 shows that the shock
configurations needed for hydrodynamic collimation are expected to
begin to operate with initial parameters and on size scales compatible
with those derived for YSOs.

\section{Results}
\subsection{Basic Flow Pattern}
In this section we focus on the results of a single jet producing
simulation: case A in Table 1.  At the end of this section we explore
the role of initial conditions on the final flow. For now we note that
while the density contrast used in case A is high ($q = 70$) the
features seen are characteristic of all the other simulations where
jets appear ($q \ge 7$).  In Figs 3 and 4 we present results of the
case A simulation after 1035~years of evolution.  Figure 3 shows a
gray scale map of the logarithm of the density alongside of a vector
map of the velocity field.  Figure 4 shows gray scale maps of both
temperature and pressure.  Note that the darkest gray tones in Fig~3
correspond to low values of the density, whereas in Fig~4 the darkest
gray tones correspond to high values of temperature and pressure.

Figures 3 and 4 demonstrate that the central wind, emerging from the
base of the grid, becomes highly focused through the interaction with
the environment.  While there are features of the overall flow pattern
that resemble a wind blown bubble, the shocked wind has clearly been
collimated into a supersonic jet.  The collimation can be most clearly
seen in the velocity vectors in Fig~3.  These show a high speed flow
above (behind) the wind shock, aligned with the $z$ axis. 

In order to understand the nature and origin of the flow pattern we
focus first on the density map.  In the parlance of wind-blown bubble
theory discussed in section 3 we can define the outer boundary of the
interaction region or ``bubble'' by the shock wave driven into the
ambient medium by the central wind.  Behind this ambient shock is a
shell of swept-up, compressed ambient gas. At any height $z$ the
highest densities in the flow are found in this shell.  The inner
boundary of the bubble is defined by the shock wave which faces into
the central wind, decelerating and heating it.  It is the mildly
aspherical feature at the base of the computation domain surrounding
the freely expanding spherical wind. 

Ignoring the flow interior to the swept-up shell for the moment, the
elongation of the bubble can, to first order, be explained in a simple
way.  Note first that the ambient pressure can play no role in shaping
the bubble.  The highest pressures in environment are achieved in the
equator where $P_{\rm e} \propto \rho_{\rm e} T_{\rm e}$. Even there
the pressure there is always orders of magnitude lower than the driving
thermal pressure achieved in the shocked wind with $P_{\rm sw} \propto
\rho_{\rm w} {v_{\rm w}}^2$.  Thus only the inertia of the environment
affects the shape of the bubble.  In his study of wind blown bubble
dynamics Icke (1988) used Kompaneets' (1960) formalism to derive an
expression for the evolution of the ambient shock,
\begin{equation}
R_s = R_s(\theta,t)
\label{rsdef}
\end{equation}  

\begin{equation}
{\partial R_{\rm s} \over \partial t} = {\lbrace A (1+({1 \over
R_{\rm s}}{\partial R_{\rm s} \over \partial \theta} )^2)\rbrace}^{1
\over 2}
\label{Komp}
\end{equation}

where
\begin{equation}  
A = {\gamma + 1 \over 2} {P_{\rm sw} \over
\rho_{01} (\theta)}
\label{accpar}
\end{equation}
Equation \ref{Komp} shows that $A = A(\theta)$ can be defined as a
local acceleration parameter for the ambient shock.  Therefore the run
of $A(\theta)$ determines the asphericity of the bubble. Since the
Kompaneets approximation assumes that $P_{\rm sw}$ is constant across
the shocked wind cavity it is the environments density distribution
(inertia) which determines the $\theta$ dependence of $A$.

The flow of shocked wind interior to the swept-up shell
departs strongly from the expectations of wind-blown bubble theory.
According to the classic theory of non-radiative wind-blown bubbles
the flow of the hot shocked wind should be subsonic at a high uniform
pressure $P_{\rm sw}$. But the density map in Fig~3 shows at least two
sharp discontinuities at heights $z = 8 \times 10^{16}$ cm and $z =
1.35 \times 10^{17}$ cm in the shocked wind cavity.  Comparison of the
velocity, temperature and pressure maps show that these features are
strong shock waves, which means that the flow in the cavity has been
accelerated to supersonic speeds.  Thus it is no longer appropriate to
interpret the dynamics in the simulations purely in terms of energy
driven wind-blown bubbles.  Instead we have a situation which is a mix
of jet propagation and bubble evolution physics.

These ``internal'' shocks in the cavity are quite consistent with theory
of supersonic jets.  It is well known that the interaction of a jet
with the surrounding medium will produce two shocks: a bow-shock
facing into the environment; and a jet-shock facing upstream into the
oncoming jet material (\cite{Nor93}).  Consideration of the pressure
map in Fig~4 demonstrates that the leading discontinuity at height $z =
1.35 \times 10^{17}$ cm can be identified as the the jet shock.
Indeed, the shock configuration is consistent with a Mach-disk as is
expected for a terminal jet shock.  The second ``internal'' shock wave
in the body of the jet at $z = 8 \times 10^{16}$ cm also has a Mach
disk configuration.  The origin of this feature is consistent with the
crossing shocks expected in the propagation of an initially over
pressured jet (\cite{Nor93}).

The bubble's ambient shock appears to double as a bow shock for the
collimated jet. This dual-identity is another manifestation of the mix
between jet and bubble dynamics.  Examination of animations, (which
can be seen at the WWW site
http://www.msi.umn.edu/Projects/twj/jetcol.html), as well as plots of
the various quantities along the axis (see Fig~5), show the region
between the jet and ambient/bow shocks to be more complex than might
be expected for a simple jet/environment interaction.  From the maps
shown in Figs 3 and 4 one can see a structure in this region that
resembles a kind of oblong plug not seen in previous jet simulations.
We note that simulations driving a jet into this kind of stratified
environment have yet to be performed (see \cite{DalPin95} for examples
of jet propagation simulations in non-constant environment). We will
return to the origin and evolution of the jet head or ``plug'' in the
next sub-section.

Another feature of the simulations expected from standard jet physics
is the cocoon of ``waste'' material shed by the jet across the sides
of Mach disk (\cite{Nor93}).  This material is first decelerated by
the jet shock and then is diverted to flow back around the side of the
jet body. In our simulations there is an additional feature associated
with the propagation of the jet and the cocoon.  Note the presence of
a relatively cool but dense tongues of shocked ambient material that
extends from the CD at $z \approx 5 \times 10^{16}$ cm into the
shocked wind cavity.  Given the cylindrical symmetry of these
simulations this feature acts like a chimney surrounding the jet and
helps to maintain its collimation.  Such chimneys have been observed
in other SFIC simulations and they appear to be an important element
of the inertial confinement collimation mechanism (\cite{Melal91},
\cite{IcBaFr92}, FN94).

To assist in identifying the basic features of the simulations in Fig
5 we present cuts along the $z$ axis of the density, velocity,
pressure and Mach number. The wind shock and ambient/bow shock can be
recognized in all variables at $z \approx 1000$ and 12500~AU
respectively.  Similarly the jet-shock and internal Mach disk are
apparent at $z \approx 6000$ and 9000~AU.  Note that the Mach number
is frame dependent and the distinction between subsonic and
supersonic flows is sensible only in the frame of a particular shock
wave.  Detailed examination of the evolution of the wind shock shows
that it progresses very slowly. Its rest frame and the frame defined
by the stationary grid are essentially identical and the Mach numbers
shown in Fig~5 are only correct for the flow between the wind shock
and the internal Mach disk.
 
In Fig~3 note first that the flow is clearly being accelerated from
subsonic to supersonic velocities (${\cal M} \approx 3)$ {\it after}
passing through the wind shock. This suggests the presence of a de
Laval nozzle. Note also that the average density in the body of the
jet, which we define to be the region between the wind shock and
jet-shock, is $<n> ~ \approx 100$ \cc ~ which is lower than the
density in the environment.  Thus our simulations are producing light
supersonic jets (i.e $\eta = \rho_{\rm e}/\rho_{\rm j} < 1$).
However, since the environmental density will continue to decline with
distance the jet will eventually become ``heavy'', ($\eta > 1$), if
the simulations were to be continued on a larger grid for a longer
time. In a real protostellar environment the jet would probably meet
the edge of the cloud before that happened and the jet would become
heavy more abruptly.  The addition of radiative cooling will remove
lateral pressure support for the jet and should allow it to collapse
to smaller widths and higher densities (\cite{RaCan89}).  Thus we
expect the jets produced in our simulations will always become heavy
at some point.  Finally, note again the complicated structure in the
``plug'', the region between the jet-shock and the ambient/bow
shock. We will return to this point in the next section.

The continuing collimating effect of the environment can be seen by
considering the opening angle of the jet.  The opening angle of a
freely expanding supersonic jet depends on its Mach number,
\begin{equation}
\phi = 2 \tan^{-1}({1 \over {\cal M}})
\label{phidef} 
\end{equation}
From equation \ref{phidef} the jet shown in Figs 3 and 4 would, 
if unconstrained,
have an opening angle of at least $40^{\rm o}$.  An opening angle of
$\phi \approx 22^{\rm o}$ appears more appropriate to the simulation,
indicating that continuing confinement by the chimney and
the swept-up shell are important in the dynamics of the jet.

\subsection{Evolution}
Our simulations demonstrate that a well collimated supersonic jet
develops from the evolution of a wind-blown bubble, the system being
an interesting mix of both wind-blown bubble and jet dynamics.  We
have already identified the ambient and wind shocks appropriate to
wind-blown bubbles and the jet shock, crossing shocks and cocoon of
waste jet gas appropriate to jets.  In order to make the evolution of
these features more explicit we present in Fig~6 the evolution of the
system through seven sequential gray scale maps of the density, taken
every 147 years.

There are a number of noteworthy features in Fig~6.  Firstly the
evolution of the wind shock: as the system evolves it becomes more and
more aspherical, prolate geometry.  Using the distance to wind shock
in the pole $R_{\rm sw}({\rm P})$ and equator $R_{\rm sw}({\rm E})$ we can define
an ellipticity parameter to describe its geometry,
\begin{equation}
e = {R_{\rm sw}({\rm E}) \over R_{\rm sw}({\rm P})}.
\label{inshckdef}
\end{equation}
Detailed examination of the wind shock shows that after 200 years of 
evolution it assumes a quasi-steady configuration with an ellipticity of 
$<e> \approx .75$.  As we will see the asphericity of the wind shock plays
an important role in the collimation of the flow (Sect 5.3).  

Capturing the wind shock poses special challenges for the numerical
code as it is strong and extends over a relatively small region.  One
of the disadvantages of using a cylindrical code is the difficulty in
modeling quasi-spherical structures on a small number of grid points.
At later times in the evolution of the models we find that numerical
errors appear in the wind shock.  These are apparent in the small
``flame'' shaped region of low density immediately behind the wind shock
and close to the symmetry axis.  By viewing animations of the
simulation we have found that this feature produces a small but
noticeable effect on the evolution of the jet. The distortion of the
wind shock drives a periodic modulation in the post-shock velocity.
The pulses can be seen in velocity plot shown in Fig~5.  Fig~6 shows
that until $t \approx 900$~years there is a crossing shock in the
jet at a distance about $z \approx .5 R_{\rm sw}({\rm P})$, which
can be explained as the expansion and subsequent contraction
of an overpressured jet (\cite{Nor93}).  The velocity pulses produced
at the wind shock however change the crossing shock into an ``internal''
Mach-disk, a structure of similar character and origin to the internal
working surfaces explored by \cite{BiRa94}.

Fig~6 also demonstrates the role of the collimating chimney.  As the
system evolves relatively cool and dense shocked ambient gas is
continually pulled off the CD.  This material is
driven upwards into the shocked wind cavity where its inertia helps
maintain the collimation of the jet. Comparison of Figs 3 and 4 shows
the correlation between pressure in the jet and the shape of the
chimney.  The kink in the chimney occurs at roughly the same height as
the crossing shock.

In order to test the sensitivity of the chimney structure to numerical
viscosity we did a series of simulations with increasing resolution.
We doubled the resolution from $64\times320$ through
$256\times1280$.  Each grid doubling reduces the numerical viscosity by
a factor of $4$.  Reducing the viscosity in this way did not effect
the existence or evolution of the chimney other than steepening the
density gradients.  Due to constraints on computational time we have
not, however, been able to continue the grid doubling and we can not
at this time say that our simulations are fully converged.

The physical origin of the chimney appears to be Kelvin-Helmholtz
instabilities at the CD. Near the base of the flow a large shear
gradient exists between the shocked wind and the shell of shocked
ambient gas as can be seen in the velocity map in Fig~3.  Detailed
inspection of animations shows that the chimney develops when
corrugations in the CD (assumed to originate from KH instabilities)
are convected up by the bulk flow in the shocked wind cavity.

Some aspects of the evolution of the ``plug'' at the head of the jet
can also be followed in Fig~6.  Consideration of this figure and
density plot in Fig~5 shows that there are two contact discontinuities
in the plug.  In Fig~5 these occur at $z \approx 9500$ and $z \approx
12000$ AU respectively.  From Fig~5 it can be seen the contact
discontinuity occurring at at $z \approx 12000$ AU is the inner edge
of the swept-up shell of ambient gas.  The contact discontinuity at $z
\approx 9500$ AU marks the contact discontinuity at inner edge of what
would be, in a classic jet, the bow shock.  In our simulations however
much the material between these two contact discontinuities originates
in the stellar wind.  From inspection of the early epochs of the
simulations it appears that the plug initially forms from subsonic
material injected into the shocked wind cavity before the jet forms.
This material entered the cavity at relatively high densities and was
then further compressed by the jet once it develops.  At later times
however Fig~3 and 6 show that additional material is added to the plug
as shocked jet gas exiting the Mach-disk ``splatters'' against the
shell and is driven both backward into the cocoon and forward into the
plug.  Ambient material appears to be pulled into the plug at these
points as well during the later evolution of the jet/bubble.

The mixture of wind-blown bubble and jet dynamics can be quantitatively
explored by examining the evolution of three characteristic lengths:
the distance to the ambient shock along the pole $R_{\rm s}({\rm P})$; the
distance to the ambient shock along the equator $R_{\rm s}({\rm E})$; the
radius of the wind shock $R_{\rm sw}$.  Recall that $R_{\rm
sw}({\rm P})\approx R_{\rm sw}({\rm E})$.  For a spherically expanding bubble both
$R_{\rm s}(t)$ and $R_{\rm sw}(t)$ have closed form analytical
expressions.
\begin{equation}
R_{\rm s}(t) = \lambda_1 \lbrack 
{{\dot M}_{\rm w} {v_{\rm w}}^2 \over \rho_{01}} \rbrack^{2 \over 7}
          t^{6 \over 7}.
\label{aash}
\end{equation}
\begin{equation}
R_{\rm sw}(t) = \lambda_2 \lbrack {{\dot M}_{\rm w} {v_{\rm w}}^{5 \over 6} \over 
\rho_{01}} \rbrack^{3 \over 7}
          t^{11 \over 14}.
\label{awsh}
\end{equation}
where both $\lambda_1$ and $\lambda_2$ are constants of order $1$.
Exact expressions for these quantities can be found in \cite{KoMc92}.
We focus first on the growth of the ambient shock.  In Fig~7 the
evolution of both $R_{\rm s}({\rm P})$ and $R_{\rm s}({\rm E})$ is plotted at
$100$~years intervals.  In addition we have also plotted the growth
expected for these shocks if $R \propto t^{6 \over 7}$. These curves
have been normalized to the time and distance of the first plotted
point.  Along the pole the ambient shock is clearly expanding faster
than the predicted $t^{6 \over 7}$ rate while long the equator it
expands slower than predicted.  The inner shock is also expanding more
slowly than its predicted rate of $t^{11 \over 14}$.  The points
plotted with an asterisk are the analytical predictions for the
magnitudes of $R_{\rm s}({\rm P})$, $R_{\rm s}({\rm E})$ and $R_{\rm sw}$
respectively at $t = 900$~years.  These values were calculated using
the appropriate mass loss rates along the equator and the pole.
Recall that the simulations of spherical bubbles recovered both the
predicted rates and magnitudes to within $10\%$. While we do not
expect spherical models to recover the magnitudes of aspherical
bubbles the growth rates should be well matched (see \cite{Dwakea95}).
From Fig~6 however it is clear that none of these quantities is
recovering the analytical growth rates for a wind blown bubble and
only $R_{\rm sw}$ is within the systematic errors of the predicted
magnitude.

Consideration of the ambient shock velocity along the pole $V_{\rm
sp}$ is also useful in determining the dynamics of the system.  The
velocity of a jet bow shock is given by the familiar formula
\begin{equation}
V_{\rm bs} = V_{\rm jet} (1 + \sqrt{\eta})^{-1}.
\label{jetbs1}
\end{equation}
If $R_{\rm s}({\rm P})$ evolved solely as a jet bow shock, then
\begin{equation}
V_{\rm sp}(R) = V_{\rm bs}(z) = 
 V_{\rm jet} (1 + \sqrt{\eta} {1 \over z^{\rm 3 \over 4}})^{-1}
\label{jetbs2}
\end{equation}
where
\begin{equation}
\eta = {\rho_{\rm 01} \over \rho_{\rm j}}.
\label{etadef}
\end{equation}
Thus the ambient shock would accelerate along the pole.
If $R_{s}({\rm P})$ evolved solely as wind blown 
bubble then
\begin{equation}
V_{\rm sp}(R) \propto \lbrack {{\dot M}_{\rm w}{v_{\rm w}}^2 
                  \over \rho_{\rm 01}} \rbrack^{1 \over 3}
          {R_{\rm s}}^{- {1 \over 6}}
\label{jetbsv}
\end{equation}
and the ambient shock would decelerate along the pole.
The lower right hand panel of Fig~7 shows the actual velocity in the
simulations as a function of radius.  For comparison we have also
plotted representative curves for both the evolution of a jet bow
shock and a wind blown bubble. Apart from small variations, the
velocity is roughly constant at $v_{\rm sp} \approx 60$
\kms.  Thus the simulations indicate that in spite of the clear
presence of a jet the global dynamics of the system lies between that
of a pressure driven bubble and supersonic jet.

\subsection{Collimation and Initial Conditions}
In order to test the limits of the hydrodynamic collimation mechanisms
under study we have explored the parameter space of initial conditions
for the simulations.  We find that the density contrast q is the most
important parameter for determining the collimation of the flow.  In
Fig~8 we show contour plots of density from four different simulations
(case C, D, E and F) each with different initial values of the equator
to pole contrast $q$ ($q = 3, 7, 14$ and $30$ respectively).  These
plots demonstrate that collimation of the shocked wind flow into a jet
occurs between $q = 7$ and $q = 14$.  These are not extreme values.
The values of $q$ obtained in numerical simulations of the collapse of
rotating clouds can be as high as 1000 (\cite{Yorket93}).  Recall also
that the collimation of the flow occurs without the benefit of
additional ram pressure from the inward directed accretion flow that
would occur with more realistic initial conditions. Since the
accretion velocity goes as $R^{\rm ({-{1 \over 2}})}$ the asphericity
which develops from the density gradient will be enhanced as the
equatorial shock will remain at relatively smaller radii where the ram
pressures are higher.  The same principal should hold for the
inclusion of gravitational potential of the protostar.

We have also performed a set of runs to determine the effect of wind
luminosity on the jet collimation.  Using the results from
\cite{Smithet83}, \cite{KoMc92} fixed the following limits on the
effectiveness of hydrodynamic collimation in terms of the wind
luminosity ($L_{\rm w} = {1 \over 2} ~{\dot M}_{\rm w}{V_{\rm w}}^2$),
\begin{equation}
3 < {L_{\rm w} \over L_{\rm wc}} < 10
\label{lumlim}
\end{equation}
In equation \ref{lumlim} $L_{\rm wc}$ is a critical luminosity based
on, among other things, the scale height of the ambient density
distribution.  Since no characteristic scale exists for a power-law
distribution a direct comparison with this prediction is difficult.
However, we can bracket the range of luminosities where the
collimation seen in our simulations operates. We have run two
simulations (case G and H) with the density contrast fixed at $q = 20$
but varying the velocity from $V_{\rm w}= 100$ \kms\ to $V_{\rm w} =
700$ \kms.  These simulations cover a factor of $49$ in the wind
luminosity, nine times larger than the range predicted by equation
\ref{lumlim}.  We find jet collimation occurs in both simulations.
Thus, given the idealizations inherent in our model we find no
significant limits on hydrodynamic collimation over a range of
luminosities consistent with with those observed in YSOs. In the next
section we discuss some reasons why equation \ref{lumlim} might be
wrong.

\section{Collimation Mechanisms}
From the results presented in the previous section it is clear that it is
possible to use the interactions with the environment to produce a
high degree of collimation in the shocked wind as well as transsonic
flow.  From these simulations and others studies of inertial
collimation it appears that a number of mechanisms contribute to the
production of supersonic jets.  In this section we explore these
mechanisms in more detail.

\subsection{De Laval Nozzles}
In Fig~9 we present a contour plot of model B after 300 years of
evolution with cuts of velocity, pressure and Mach number along the
pole plotted beneath it.  As in Fig~5 the axial cuts show that
immediately behind the wind shock, $R_{\rm sw}({\rm P})$, the flow passes
through a sonic point. We have also marked the height at which the
contact discontinuity achieves a minimum width (ie $\min[r_{\rm
CD}(z)]$) in the plots shown in Fig~8. It is clear that the
constriction in the walls of the shocked wind cavity occurs at the
same point where the flow undergoes a sonic transition, in other words
a de Laval nozzle has formed.

The presence of de Laval nozzles is not, in itself, surprising.  As
was noted in the introduction there is an extensive literature on de
Laval nozzles as jet collimation mechanisms.  Currently it appears
that this type of mechanism is out of favor.  But the manifestation
of de Laval nozzles seen in these simulations is quite different from
the standard steady state models found in the literature.  Therefore
the conclusions which led to the abandonment of de Laval nozzles do
not apply here.

As was discussed in the previous subsection \cite{Smithet83} placed
stringent limits on the parameter space of initial conditions under
which stable de Laval nozzles were expected to form (eq \ref{lumlim}).
The theoretical basis for this conclusion was the expectation that
beyond these limits Kelvin-Helmholtz instabilities would choke off the
nozzle producing a series of bubbles rather than a continuous jet.
The analytical arguments invoked were bolstered by the results of
numerical simulations carried out in an earlier paper by
\cite{Normanet81}.  Comparing these papers with our results is
difficult because of the different initial conditions and assumptions.
In spite of these differences, however, a few points can be made.

Firstly, and most obvious, the numerical simulations carried out by
\cite{Normanet81} are highly underresolved by current standards
(though they were state of the art at the time).  The simulations were
performed on $40 \times 40$ computational grids, which is almost a
factor of 64 smaller than used in the simulations presented here.
Also, one can analytically show that Kelvin-Helmholtz instabilities
occur, but not that they will choke off the jet nozzle. What matters
is the non-linear, time-dependent effect of the instabilities on the
flow.  We see in our simulations that the strong shear along the
contact discontinuity does produce ripples along its surface, but
rather than choke off the jet we find the instabilities end up aiding
its collimation by providing the material for the dense chimney.

There is another important difference between our simulations and
previous studies of de Laval nozzles.  For the most part analytical
investigations have assumed a steady state configuration for the
nozzle by matching the pressure in the bubble with the pressure in the
ambient medium (see \cite{Kon82} for some thoughts on the evolutionary
aspects of nozzles).  In our model the bubble is over pressured and
keeps expanding.  Here it is the inertia, not the pressure of the
ambient medium which provides the basis for producing a collimating
cavity.  This means that if our models are to be applied to YSOs the
wind or the accretion or both must be taken to be time dependent
allowing the configuration to maintain a stable average configuration.
There is considerable observational support for such a conclusion as
the jets themselves are known to be time dependent (\cite{Mors92}) We
will take up this issue again in the final section.

Finally we note the issue of radiative time scales. De Laval nozzle
models require the presence of high temperature gas.  This has led to
the conclusion that the size scales required, i.e.\ those of adiabatic
bubbles, are too large to allow nozzles to collimate on the scales
observed.  However, recognition of the PRB phase discussed in Sect 3
allows de Laval nozzles to operate on considerably smaller scales than
was previously thought (see also \cite{RaCan89}).  This implies that
de Laval nozzles may still be an important aspect of hydrodynamic
collimation mechanisms for YSO jets.

\subsection{Shock Focusing}
While de Laval Nozzles are clearly an important aspect of the
collimation of jets in these simulations there is another mechanism at
work.  As was noted earlier the wind shock is not spherical.  In
almost all the simulations which produce jets we have found wind shock
ellipticities of $0.7 < e < 0.8$ (see eq \ref{inshckdef}) In addition
simulations of similar astrophysical systems (i.e. Planetary Nebulae,
SN1987A, Superbubbles)
have found stronger departures from spherical geometries with
ellipticities achieved as low as $e = 0.25$ (\cite{Icke94},
\cite{Dwakea95}).  When the wind shock takes on prolate geometries the
radially streaming wind from the central source encounters it at an
oblique angle.  Only the normal component of the wind velocity will be
shocked in these cases.  The tangential component will remain
unchanged. Thus the wind shock can act as a lens focusing the
post-shock velocity vectors towards the jet axis.

In Fig~10 we present a map of the post shock velocity vectors
overlayed on contours of density to show the position of the inner
shock and CD.  In this figure the shock has a ellipticity of $e = .79$.
Note that the flow vectors close to the equator emerge from the shock
without any focusing.  These gas parcels are only turned towards the
axis by pressure gradients at some distance downstream.  But as one
travels up towards the pole the flow vectors are clearly being
refracted poleward {\it directly behind the shock}.  As we shall
demonstrate below flow focusing by the wind shock can be an important
component of the collimation process containing the potential to
produce fully supersonic jets without the presence of a de Laval
nozzle.

\subsection{Properties of the Inner Shock}   
It is difficult to make an a priori determination of the wind shock
shape based only on initial conditions. The degree to which the wind
shock departs from spherical symmetry will be determined by nonlinear
feedback, in terms of both thermal and ram pressures, from the
evolving bubble. It is not yet clear how to calculate the
characteristics of this feedback analytically.  Because of this
difficulty almost all analytical treatments of aspherical wind-blown
bubble evolution have assumed the inner shock to be spherical (see
e.g.\ \cite{Smithet83}, \cite{MacMc88}).  In contrast almost every
study of these systems relying on numerical simulations has shown
the inner shock to be aspherical to some degree (\cite{MacMcN88},
\cite{BloLun93}).

Assume the shock takes on an elliptical geometry with ellipticity $e$
defined by eq \ref{inshckdef}. To determine the degree of focusing in
the post wind shock flow we must solve the oblique shock jump
conditions.  Here we repeat and extend the analysis of Icke (1988).
Working in spherical coordinates, the angle ($\beta$) between the
ellipse and radially directed wind will be a function of polar angle
($\theta$) and is given by
\begin{equation}
\beta=\theta+\arctan\left({ e^2 \over \tan{\theta}}\right)\;,
\label{betdef}
\end{equation}
Because the wind shock is aspherical the radial distance at which the
wind will encounter the shock depends on latitude.  Thus the
geometrical dilution of the wind will cause the pre-shock flow
variables to be functions of the polar angle. We denote the pre-shock
variables with the subscript ``0''.  Accounting for these variations we
can use the jump conditions for a strong shock to express the
post-shock variables (denoted with subscript ``1'') as
\begin{equation}
  v_{\rm 1p}(\theta)=\left({\gamma-1 \over \gamma+1}\right)v_0
\sin\beta(\theta),
\label{vvar1}
\end{equation}

\begin{equation}
\qquad v_{\rm 1t}(\theta)=v_0\cos\beta(\theta)
\label{vvar2}
\end{equation}

\begin{equation}
P_1(\theta)=P_0 \left({2 \over \gamma+1}\right){\cal M}_0^2(\theta)\sin^2(\beta(\theta))
\label{pvar}
\end{equation}

\begin{equation}
s_1(\theta)=\left({\gamma-1 \over \gamma+1}\right)s_0(\theta)
\sqrt{{2\gamma \over \gamma-1}{\cal M}_0^2(\theta)\sin^2\beta(\theta) -1}\;.
\label{svar}
\end{equation}

In Eqs. \ref{vvar1} through \ref{svar}, $v_{\rm 1p}$ and $v_{\rm 1t}$
are the post-shock velocity parallel and tangential to the shock
normal, $s$ is the sound speed and ${\cal M}$ is the Mach number.  The
total angle of defection ($\chi$) through the shock is given by
\begin{equation}
  \tan\chi(\theta)={2\tan\beta(\theta) \over
(\gamma+1)+(\gamma-1)\tan\beta(\theta)}\;.
\label{chivar}
\end{equation}

In Fig~11 we show curves of $\chi$, $P_1$, speed $v_1 = \sqrt{v_{\rm
1p}+ v_{\rm 1t}}$, and post-shock Mach number ${\cal M}_1$ as a
function of polar angle for three values of the shock ellipticity $e$.
Note that all points along the inner shock where $\chi > \theta$ have
fully focused post-shock velocity vectors i.e. the flow at these
points is directed towards the polar axis.  For the three values of
$e$ shown all wind streamlines within $\theta_{\rm f} = 30^{\rm o}$ of
the polar axis exit the shock fully focused.  Here we define
$\theta_{\rm f}$ to be the latitude below which the flow is fully
focused (i.e. $\chi = \theta$). In our simulations we found $<e>
\approx .75$ which yields $\theta_{\rm f} = 33^{\rm o}$. Thus, as Icke
expressed in his original study, ``even a small eccentricity causes a
high degree of focusing''.

The plots of velocity demonstrate the extent to which the wind can
pass through the prolate inner shock and emerge into the bubble
without being significantly decelerated.  In some sense a highly
prolate inner shock acts much like the reconfinement shocks explored
by \cite{Sanders83} in the context of free extragalactic jets.  The
plots of pressure demonstrate that additional collimation can be
expected to occur as velocity vectors are turned poleward by
tangential pressure gradients as is also seen in Fig~10.

The plots of the post shock Mach number ${\cal M}_1$ show that for
more prolate shocks a significant region of the post-shock wind enters
the bubble with supersonic velocities. Thus in principle it is in
possible to produce a fully collimated supersonic jet purely through
the action of shock focusing. The post shock Mach number depends upon
the ratio of specific heats $\gamma$.  Since we expect that shock
focusing will begin during the Partially Radiative Bubble phase where
the polytropic index $\gamma < {5 \over 3}$ we have, in Fig~12,
plotted $\max[M_1(\theta)]$ vs $\gamma$ for 3 different ellipticities
We have chosen fairly modest values of the ellipticity ($e = .5, .65$,
and $.8$) to emphasize that supersonic flows can be expected behind a
wide range of inner shock configurations. Fig~11 shows that as
$\gamma$ decreases supersonic post shock flow can be achieved at
relatively low ellipticity. In this case no de Laval nozzles are
necessary to produce supersonic outflows.

\section{Conclusions}
The conclusions reached in this paper can be summarized as follows.
\begin{itemize}

\item The SFIC mechanism can, in principle, produce 
well collimated supersonic jets in the context of YSOs through
purely hydrodynamic means.

\item The SFIC mechanism requires the formation of a hot shocked
bubble of gas. By considering analytical estimates of the interaction
between a fast wind and its surroundings it turns out that for typical
YSO parameters this condition will hold (for outflow velocities
$\gtrsim 200$ \kms). Although the bubble will not be fully adiabatic
on small scales ($\sim 100$ AU), cooling will not be efficient enough
to stop the build up of a reservoir of hot gas. This partially
radiative phase is quite important in the YSO case and may be in other
astrophysical circumstances. A closer investigation of it would be
interesting.

\item The flow pattern that forms in the SFIC situation is a complex
mix between wind blown bubble physics and jet physics. While the
base of the jet is a reservoir of hot, subsonic gas, typical of a
wind-blown bubble, the jet itself is supersonic, showing internal
shocks, and a cocoon of ``waste'' material. It also has the
characteristic jet and bow shocks, in which the bow shock in fact is
identical to the outer shock of the bubble. The jet is over
pressured and less dense than its environment, but since the
environment does not have a constant density, the jet is expected to
change to a dense one once it has moved out to larger distances.
Also in its evolution the structure behaves in between what is
expected from analytical estimates for bubbles and jets.

\item The actual collimation in the SFIC mechanism is caused by a
combination of effects. Firstly, the interaction with the surrounding
medium creates a de Laval nozzle which allows a smooth transition from
subsonic to supersonic flow. This nozzle is evolving and is a stable
feature for a wide range of parameters in our simulations. We find
none of the unstable behavior that was previously reported for these
configurations. Secondly, the wind shock is aspherical which in itself
leads to focusing of the outflow towards the axis. In fact, an
aspherical shock may even produce super-sonic post-shock gas, since
only the normal component of the velocity is shocked. For $\gamma <
5/3$ this can happen even for mildly aspherical inner shocks.
\end{itemize}

We again note that the models explored in this paper
rely on the presence of an envelope of dense circum-protostellar
material to produce collimated jets. Thus we feel
that our results may be most relevent to the
more deeply embedded class 0 objects rather than to the more evolved
objects with disks such as T Tauri stars.

One may wonder about the long term evolution of these jets. We already
pointed out that we expect the jets to become denser than their
environment as they move out. We also pointed out that they are over
pressured. This means that the whole bubble structure expands, also
laterally and that given enough time, all of the circumstellar
material will be removed. For the region of parameter space explored
here the time scale for this is $\sim 10^4$ years. Given the dynamical
age of some HH objects it is clear that something else must occur if
the mechanisms discussed here are responsible for collimating the
jets. However, it is also clear that the jets are variable in time and
one can envisage a situation where the outflow is temporarily stopped
or weakened and the ram pressure of the accreting medium is high
enough to temporarily reverse the expansion of the bubble. The jet
production would, therefore, also stop. This would lead to a situation
similar to the one studied by \cite{BiRa94} in their numerical work on
jets from time-dependent sources.  Analytical estimates indicate that
a periodic scenario of this type does indeed work (\cite{MelFr96b}).

In the SFIC model it is unavoidable that a reservoir of hot gas forms
at the base of the jet, or even that the lower part of the jet
consists of high temperature gas. This material would in principle
emit free-free emission, observable in the radio continuum and (soft)
X-rays. However in the deeply embedded sources we are considering
these X-rays might not be observable. A specific comparison with
observations however is premature until more realistic physics is
added.  In particular the size and physical conditions in the hot
cavity will strongly depend on radiative cooling thus we defer
specific comparisons until these calculations have been carried out.

It is obvious that a substantial amount of work needs to be done
before the SFIC mechanism can be claimed to be able to explain jets
from YSOs. However given its efficiency and the fact that something
like an interaction between in and outflows must take place around
YSOs, we plan to explore it in some more detail in future
papers. Especially the effects of cooling and the partially radiative
bubble configuration will be studied in a next paper.

\centerline{{\bf Acknowledgments}}
We wish to thank Vincent Icke, Bruce Balick, Tom Jones, Arieh K\"onigl, and
Chris McKee for the very useful and enlightening discussions on this
topic. We would also like to thank Dongsu Ryu for his generous help
in making his TVD code available to us for this study.
Support for this work was provided
by NASA grant HS-01070.01-94A
from the Space Telescope Science Institute, which is operated by
AURA Inc under NASA contract NASA-26555.  Additional support came from the
Minnesota Supercomputer Institute.

\clearpage

\begin {table}[hbta]
\caption {Initial Conditions For Runs A - H}

\begin {center}
\begin {tabular} {lllllll} \hline
{run}  & {${\dot M}_{\rm w}$} & {$V_{w}$} & {${\dot M}_{\rm a}$} 
& {$q$} & {$Resolution$} \\
\hline
A  &  $1 \times 10^{-7}$ & 200  & $1 \times 10^{-5}$ & 70 & 
$256 \times 1280$ \\

B  &  $2 \times 10^{-7}$ & 300 & $1 \times 10^{-5}$ & 70 & 
$256 \times 1280$ \\

C  &  $1 \times 10^{-7}$ & 200 & $5 \times 10^{-6}$ &  3 & 
$256 \times 512$ \\

D  &  $1 \times 10^{-7}$ & 200 & $5 \times 10^{-6}$ &  7 & 
$256 \times 512$ \\

E  &  $1 \times 10^{-7}$ & 200 & $5 \times 10^{-6}$ & 14 & 
$256 \times 512$ \\

F  &  $1 \times 10^{-7}$ & 200 & $5 \times 10^{-6}$ & 30 & 
$256 \times 512$ \\

G  &  $1 \times 10^{-7}$ & 100 & $1 \times 10^{-5}$ & 20 & 
$256 \times 512$ \\

H  &  $1 \times 10^{-7}$ & 700 & $1 \times 10^{-5}$ & 20 & 
$256 \times 512$ \\

\end {tabular}
\end {center}
\end {table}

\clearpage

\begin{center}
{\bf FIGURE CAPTIONS}
\end{center}
\begin{description}

\item[Fig.~1]
{Initial density distribution.  Shown are the $\log_{\rm 10}$ contours
of density from eq \ref{rhoin} with an equator to pole contrast $q = 70$.
The two solid lines show the angle at which $\rho = .5 ~\rho_{\rm max} = 
.5 ~\rho(90^{\rm o})$.  These occur at $\theta \approx 45^{\rm o}$
making the opening angle of the density distribution $\approx 90^{\rm o}$}

\item[Fig.~2]
{Transition radii as a function of velocity.  Shown are the radii at which a
radiative momentum driven bubble makes the transition to a partially
radiative bubble (solid line) and an adiabatic bubble (dashed line).  The
curves are plotted for ${\dot M}_{\rm w} = 10^{\rm -7}$ \mdot, 
${\dot M}_{\rm a} = 10^{\rm -6}$ \mdot\ and $M = 1$ \msol.}

\item[Fig.~3]
{Density and Velocity for Model A.  Shown are a gray scale map of
$\log_{\rm 10} (\rho)$ and a map of velocity vector field for model A after
1035 years of evolution.  In the density map dark (light) shades correspond 
to low (high) densities.  In the velocity field map vectors in the inner,
freely expanding wind zone have not been plotted.  Thus the first ``shell'' 
of vectors maps out the wind shock.  }

\item[Fig.~4]
{Temperature and Pressure for Model A.  Shown are a gray scale map of
$T$ and $P$ for model A after
1035 years of evolution.  In the both maps dark (light) shades correspond 
to high (low) values.  This is the reverse of the density gray scale map 
shown in Fig~3.}

\item[Fig.~5]
{Axial plots for Model A.  Shown are plots of $\log_{\rm 10} (\rho)$,
velocity $\sqrt{{v_{\rm r}}^{\rm 2} + {v_{\rm z}}^{\rm 2}}$, pressure $P$ and 
Mach number ${\cal M}$ as a function of height above the equator $z$. All plots
are taken after 1035 years of evolution. Each plot is an
average over the first 10 zones in cylindrical radius $r$.}

\item[Fig.~6]
{Evolution of density for Model A.  Shown are seven gray scale maps of
$\log_{\rm 10} (\rho)$  for model A spaced 147 years apart. In the map dark (light) 
shades correspond to low (high) densities.  Size scales are same as that
shown in Figs~3 and 4. }

\item[Fig.~7] {Polar and equatorial shock evolution.  In left and
right upper panels the distance to the polar and equatorial ambient
shocks obtained in the simulations is shown (triangles) for 9 separate
times.  Also shown (dashed line) are curves representing the growth
predicted by analytical models of spherical self-similar bubbles.  The
point marked with an asterisk represents the predicted magnitude of a
spherical bubble with identical input conditions as the simulation
(along pole or equator).  In the lower left panel an identical plot is
shown for the wind shock along the equator.  In the lower right hand
panel the velocity of the polar ambient shock is presented
(triangles).  The dashed line represents the velocity predicted for a
wind blown bubble (normalized to the first data point).  The solid
line line represents the velocity evolution for a jet.  Note that the
solid line can not be properly normalized to the data but its form is
representative of the shape of deceleration given in eq \ref{jetbs2}

\item[Fig.~8]
{Jet collimation and equator to pole density contrast.  Shown are
$\log_{\rm 10} (\rho)$ contour plots of four models which differ only in
the values of equator to pole density contrast $q$.  Upper left: Model C
$q = 3$. Upper right: Model D $q = 7.$ Lower left: Model E
$q = 14.$ Lower right: Model F $q = 3$.}

\item[Fig.~9] {De Laval Nozzles and Outflow Collimation. In the upper
panel a $\log_{\rm 10} (\rho)$ contour map of model B after 750 years
of evolution is shown.  Below that are cuts are along of the $z$ axis
of velocity, pressure and Mach number.  The points marked on each
axial plot identify the region where the width of the channel (as
measured by the contact discontinuity) has a minimum. This is
the ``throat'' of the nozzle.}

\item[Fig.~10]
{Shock Focusing: Model A.  Shown are selected $\log_{\rm 10} \rho$ contours
identifying the wind shock and contact discontinuity.  Also shown are velocity
vectors for computational zones immediately downs stream of the wind shock
($v < v_{\rm w}$}).  The density contours are $\log_{\rm 10} \rho = 
[-20.95,-20.9,-20.85,-20.2,-20.0,-19.8,-19.6,-19.2,-19.0]$}

\item[Fig.~11]
{Shock Focusing: Analytical Model.  Post-shock flow variables as a function
of polar angle for 3 elliptical (prolate) shocks of differing ellipticity.
These plots are for a wind velocity of 250 \kms\ and a wind density
at the equator of 200 \cc.
Upper left: total deflection angle. Upper right: Mach Number ${\cal M}$.
Lower left: Gas Pressure $P$. Upper right: Velocity $v$.  The ellipticities
of the shocks are $e = .3$ (dotted line), $e = .5$ (dashed line), $e = .8$
(dash-dot line).  In the plot of total deflection angle the solid line 
corresponds to $\chi = \theta$.  All points to the left of this line have
post-shock velocity vectors that are fully focused, i.e. they point
towards the z axis.}

\item[Fig.~12]
{Post-Shock Mach Numbers for Non-Adiabatic Wind Shocks. Shown are the maximum
value of the post-shock Mach number for 3 shocks of differing ellipticity
as a function of the polytropic index $\gamma$. The ellipticities
of the shocks are $e = .8$ (dotted line), $e = .65$ (dashed line), 
$e = .5$
(dash-dot line).} 

\end{description}

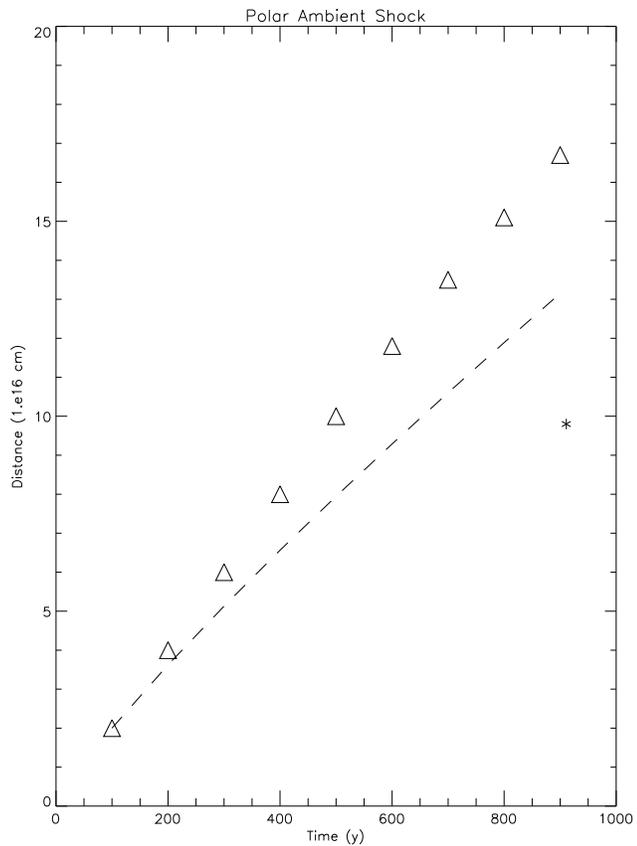
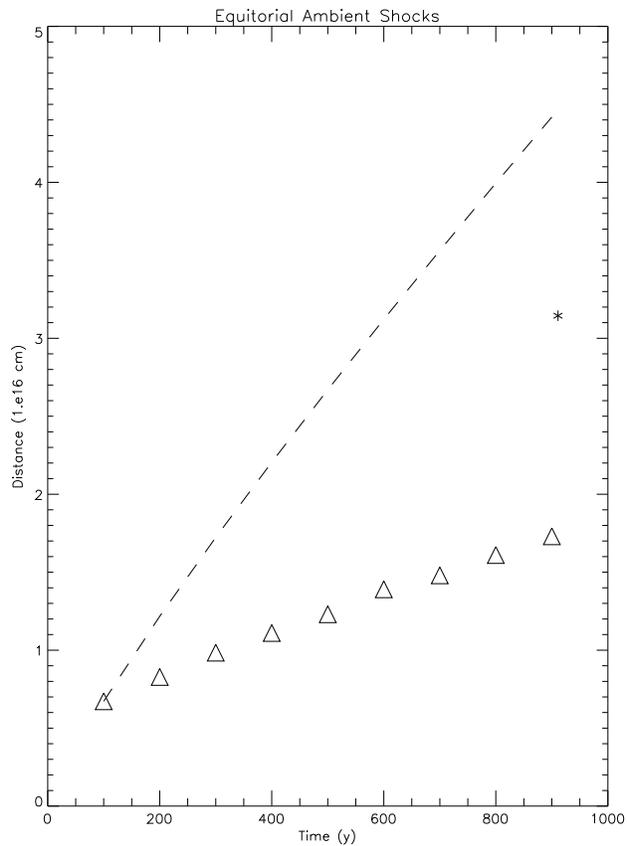
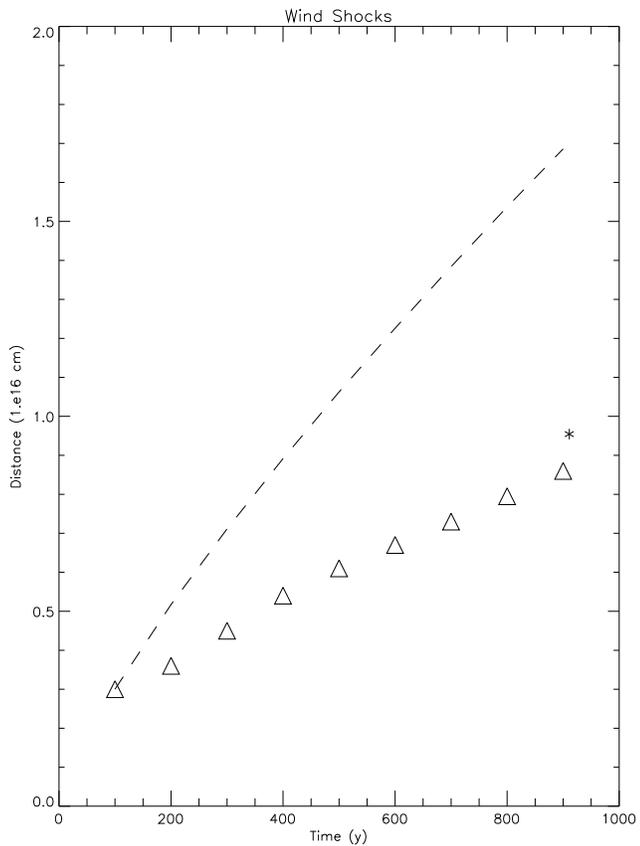
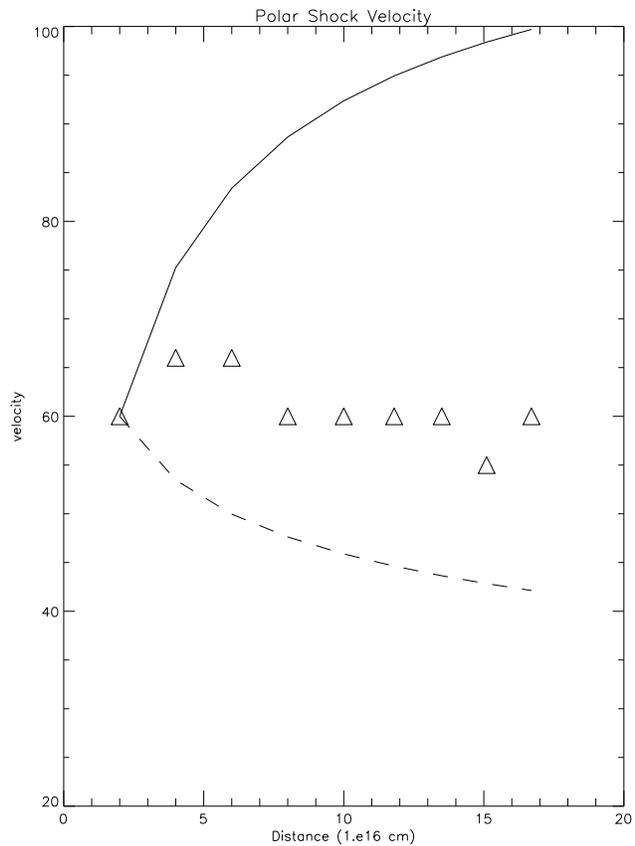
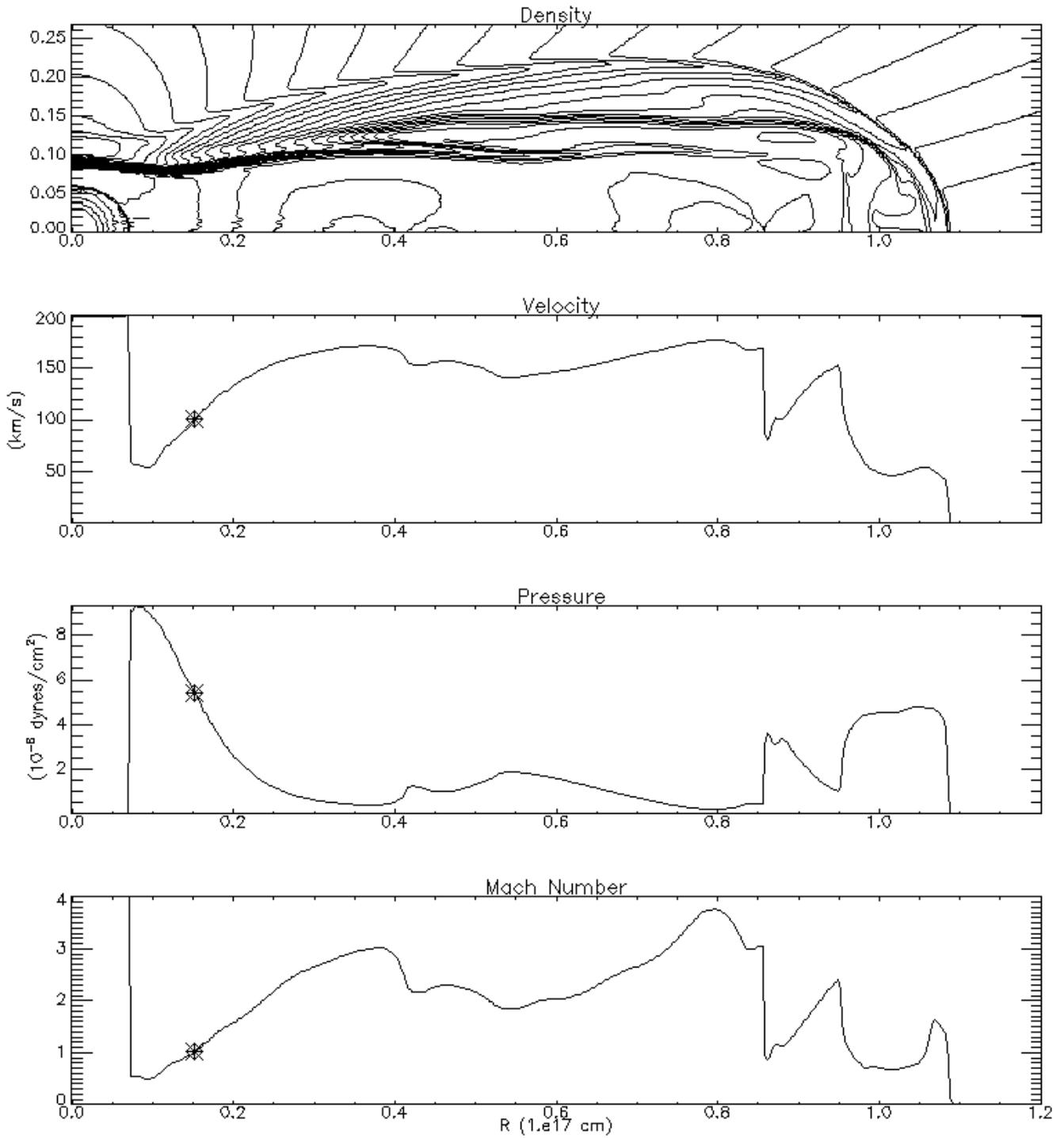
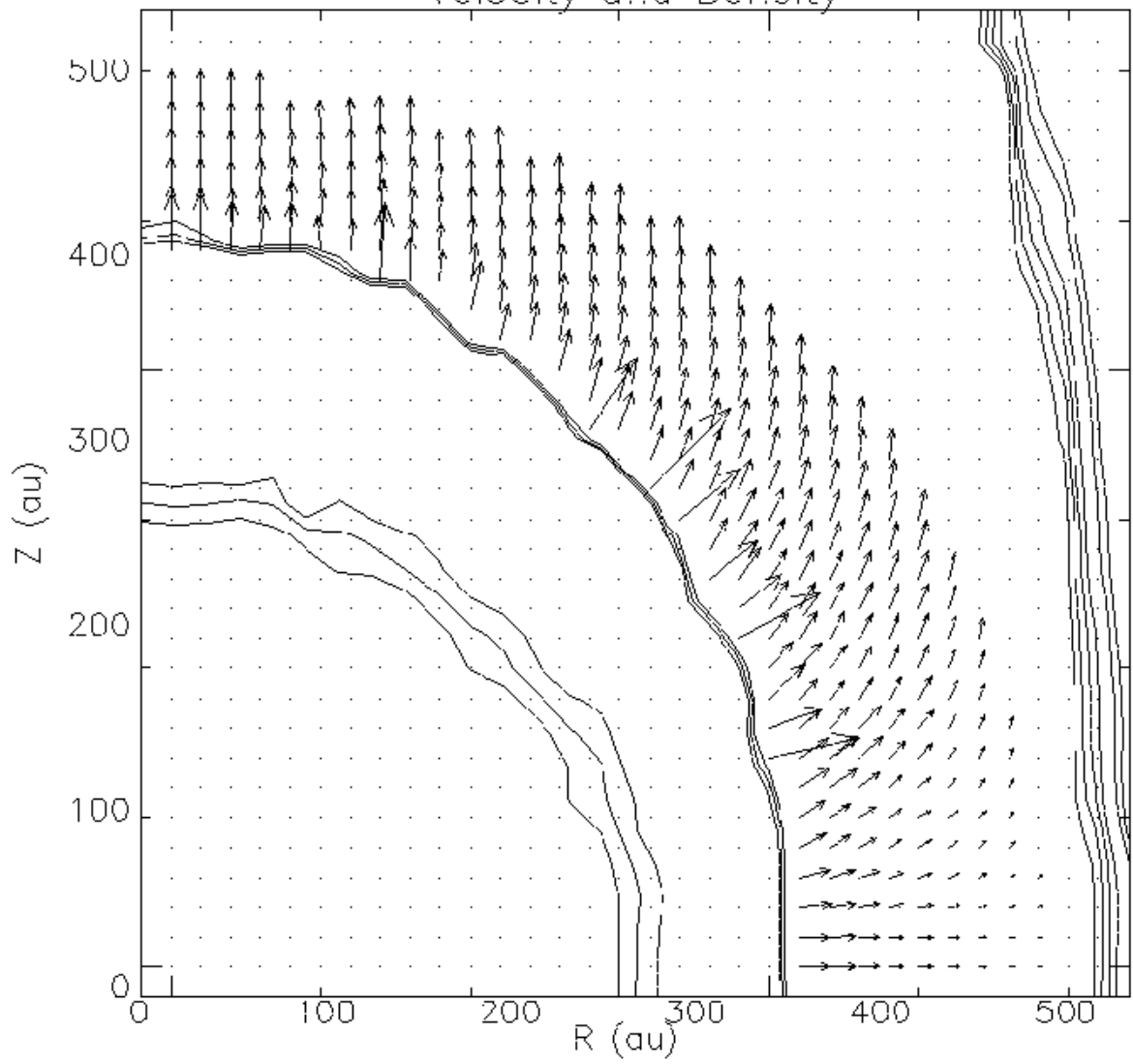
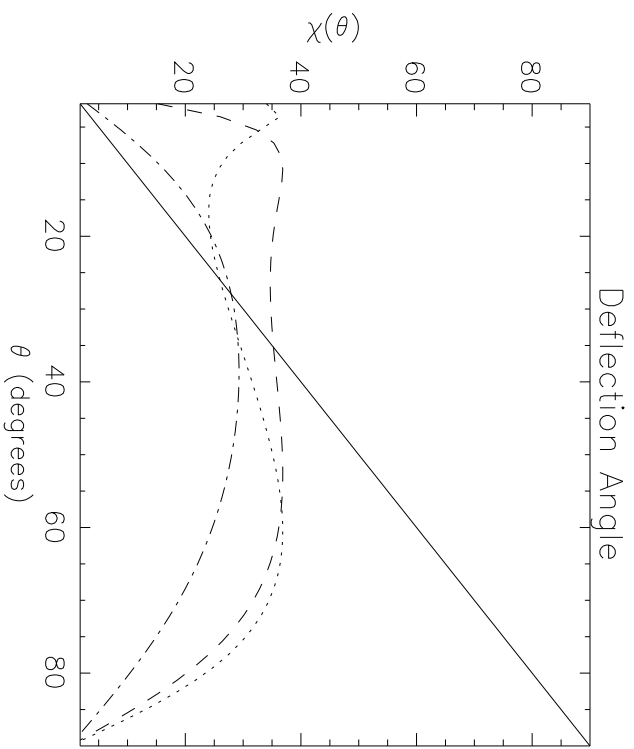
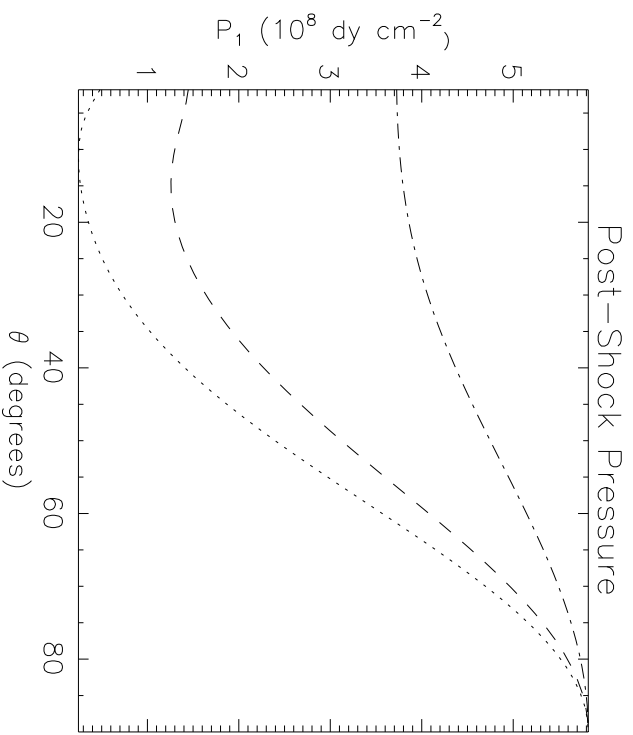
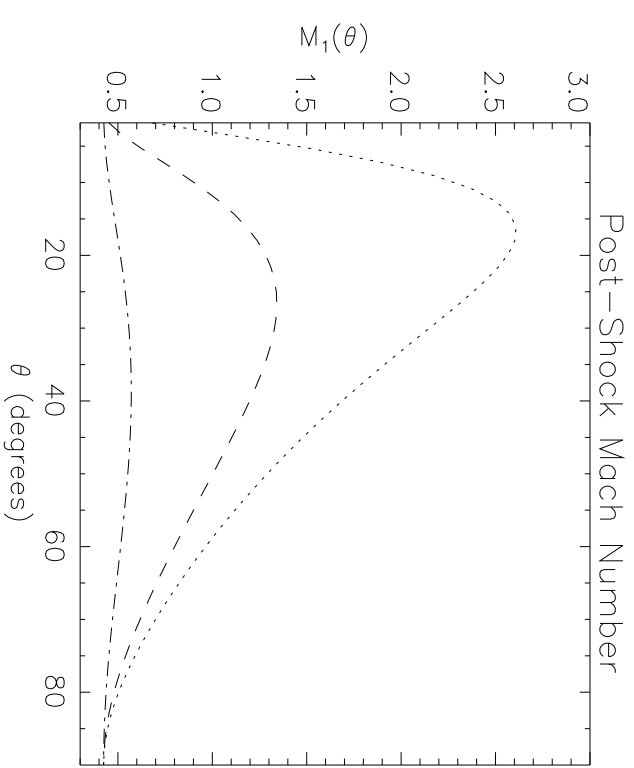
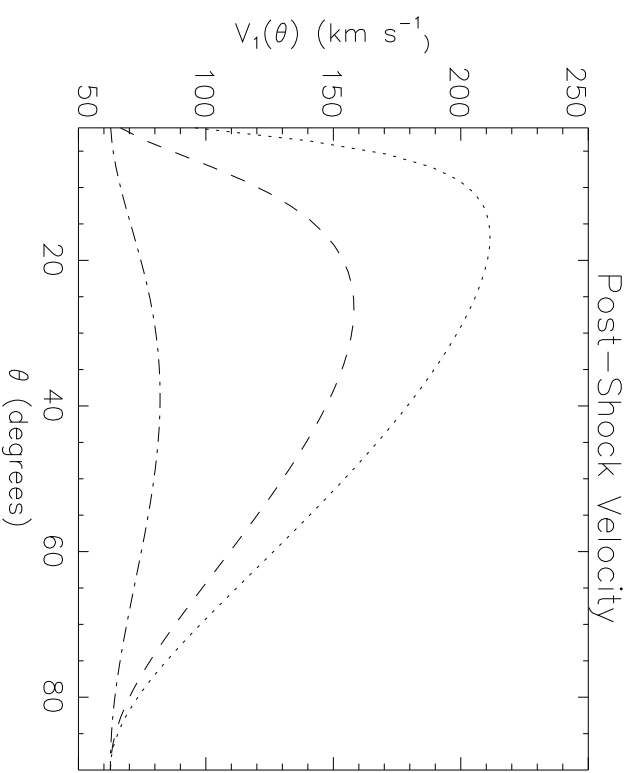
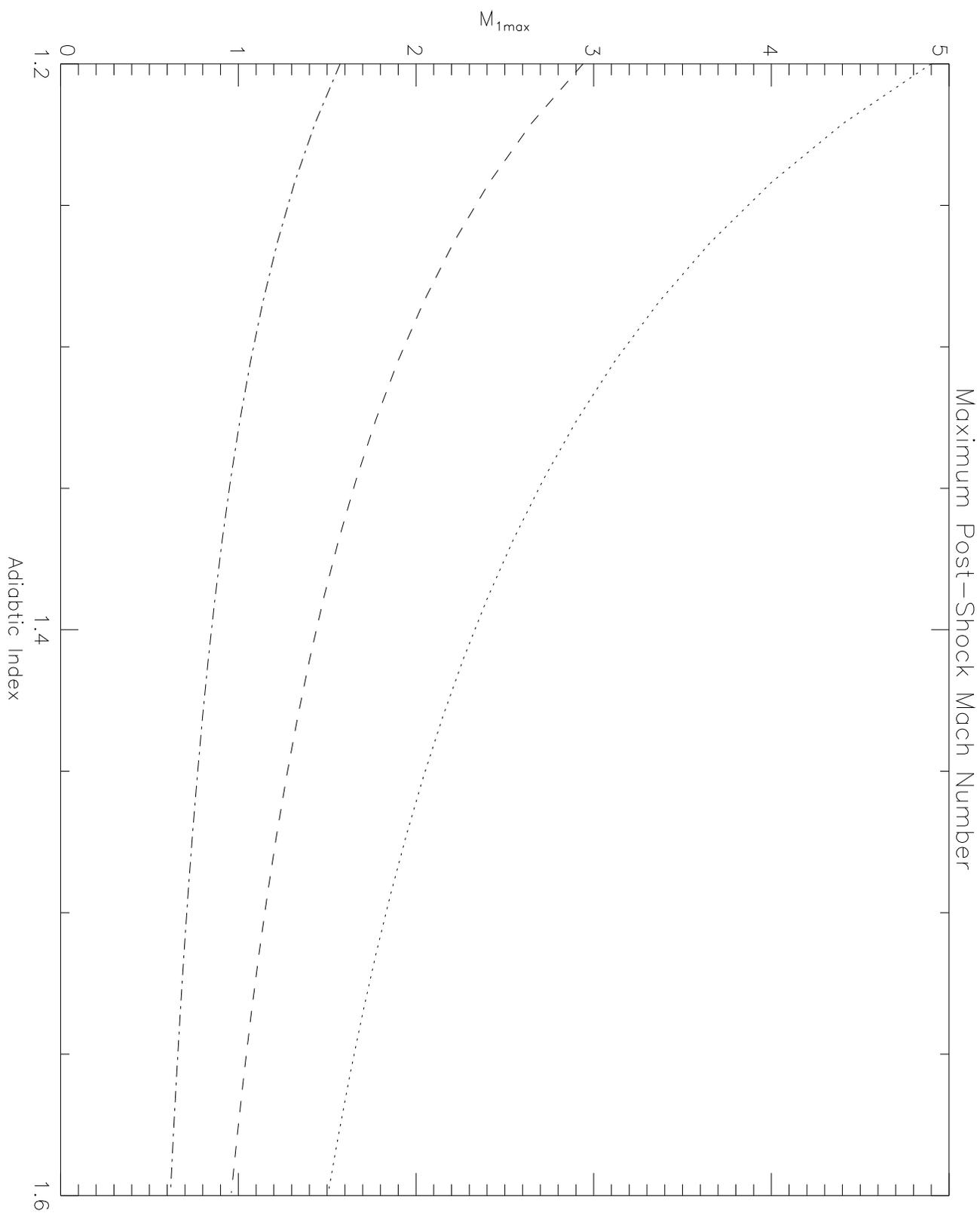
\end{document}